\documentclass{aa}
\usepackage{psfig}
\usepackage{natbib}
\bibpunct{(}{)}{;}{a}{}{,}

\begin{document}

\def\eps{\varepsilon}
\def\araa{ARAA}
\def\apss{APSS}
\def\aap{A\&A}
\def\apj{ApJ}
\def\apjl{ApJL}
\def\mnras{MNRAS}
\def\aj{AJ}
\def\prl{RRL}
\def\nat{Nature}
\def\aaps{A\&A Supp.}
\def\physrep{Phys. Rep.}

\def\me{m_\e}
\def\lesssim{\mathrel{\hbox{\rlap{\hbox{\lower4pt\hbox{$\sim$}}}\hbox{$<$}}}}
\def\gtrsim{\mathrel{\hbox{\rlap{\hbox{\lower4pt\hbox{$\sim$}}}\hbox{$>$}}}}

\def\h{h_{50}}

\def\del#1{{}}
%\del#1{{#1}}

\title{Modification of cluster radio halo appearance by the thermal
Sunyaev-Zeldovich effect} 
\titlerunning{Modification of cluster radio halo appearance by the thermal
SZ-effect} \author{Torsten A. En{\ss}lin}
\authorrunning{T. A. En{\ss}lin} \institute{Max-Planck-Institut
f\"{u}r Astrophysik, Karl-Schwarzschild-Str.1, Postfach 1317, 85741
Garching, Germany} \date{Received 11.10.2002 / Accepted 31.10.2002}

\abstract{We discuss the consequences of the spectral and
morphological modification of galaxy cluster radio halos due to the
Sunyaev-Zeldovich (SZ) effect for the interpretation of existing and
upcoming high frequency radio observations.  Likely these
modifications have affected the interpretation of the existing Coma
cluster radio data. The radio halo emission visible at low ($\lesssim
5$ GHz) frequencies is at higher ($> 10$ GHz) frequencies completely
over-compensated by the thermal SZ decrement. Thus, the total radio
emission of a galaxy cluster goes through zero (in comparison to the
constant cosmic microwave background (CMB) emission) at a frequency of
several GHz. Since the radio halo brightness has a narrow radial
profile compared to the SZ decrement, a central emission region is
surrounded by a decrement within the intermediate frequency range of
several GHz. The size of this emission regions shrinks with increasing
frequency until the decrement dominates everywhere in the cluster.
 } \maketitle

\section{Introduction\label{sec:intro}}

Clusters of galaxies are filled by a hot ($kT_{\rm e} \sim 5$ keV)
plasma. Some -- if not all -- clusters host ultra-relativistic
particle populations, observable by synchrotron emission of $\sim 10$
GeV electrons. This emission forms the {\it cluster radio halos}.
Recently, the number of detected cluster radio halos increased
significantly \citep{1999NewA....4..141G, 2000NewA....5..335G,
2001ApJ...548..639K} and the prospects for upcoming sensitive radio
telescopes to detect large numbers of cluster radio halos are
promising \citep{astro-ph/0209218}.

Radio halos seem to be triggered by strong cluster merger
\citep{2001ApJ...553L..15B}, although the detailed physical mechanism
providing the energetic electrons is still to be revealed. Two classes
of theoretical models are discussed in the literature: i) in-situ
acceleration of the rapidly cooling electrons by turbulence and shock
waves throughout the cluster volume \citep[][ and
others]{1977ApJ...212....1J}, and ii) secondary electron injection
after particle physical interactions.
In-situ acceleration typically predicts steepening radio spectra due
to the decreasing ratio of acceleration efficiency to radiative
losses for higher energy electrons. In order to confirm such models a
detailed knowledge of the high frequency radio halo spectrum and
morphology is desirable.
The classical secondary injection mechanism is the decay of charged
pions which were produced in hadronic interaction of relativistic
protons with the background gas \citep[][ and
others]{1980ApJ...239L..93D}. However, also the self-annihilation of
neutralino dark matter particles was proposed as an electron source
\citep{2001ApJ...562...24C} although the observed correlation of radio
halos with merging events clearly disfavours this
scenario. Nevertheless, such dark matter based electron sources may
lead to a lower level of radio emission and a search for such emission
has the potential to constrain scenarios of the nature of the dark
matter \citep{astro-ph/0208458}.

Such desirable high frequency measurements of the cluster halo
emission will unavoidably be affected by the thermal SZ effect
\citep{1969Ap&SS...4..301Z, 1980ARA&A..18..537S} \del{,
1999PhR...310...97B,2002ARA&A..40..643C} of the hot electrons in the
cluster plasma, as was also pointed out by
\cite{2000ApJ...544..686L}. Our goal is to discuss the effects arising
from the interference of these two radiation mechanisms.

We use $H_0 = 50\,\h\,{\rm km\,s^{-1}\,Mpc^{-1}}$, and assume $H_0 =
65\,{\rm km\,s^{-1}\,Mpc^{-1}}$ in numerical examples.

\begin{figure}[t]
\begin{center}
\psfig{figure=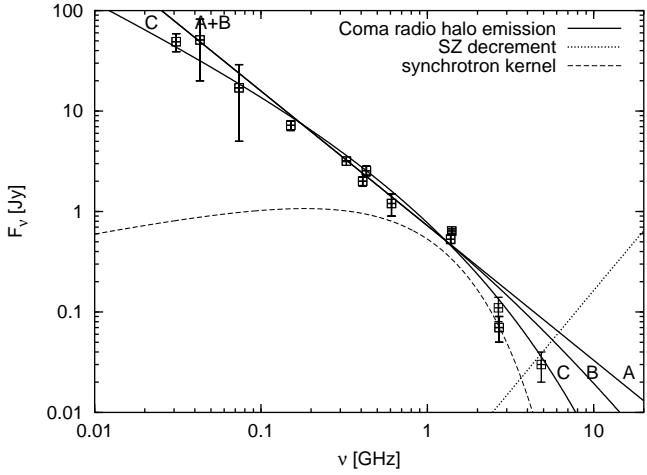,width=0.5 \textwidth,angle=0}
\end{center}
\vspace{-0.8cm}
\caption[]{\label{fig:spec1} Observed radio halo fluxes of the Coma
cluster as compiled by \cite{astro-ph/0210147}. The three radio halo
emission models (A,B,C) described by Eqs. \ref{eq:mod1}-\ref{eq:mod3}
are displayed in comparison to the isotropised synchrotron kernel (see
text). The SZ decrement (a negative flux) is also shown. The solid
angle over which the observed fluxes were integrated may vary from
observation to observation. For the theoretical models, the flux from
the full cluster volume up to the assumed position of the accretion
shock is given.}
\end{figure}

\section{Thermal Sunyaev-Zeldovich effect}

The CMB has a perfect blackbody brightness spectrum
%\begin{equation}
$I_{\rm cmb}(\nu) = i_0\,i(x),$
%\end{equation}
with $x = h\,\nu/(kT_{\rm cmb})$, $i_0 = 2\, (kT_{\rm
cmb})^3/(h\,c)^2$, and $i(x) = x^3/(e^x -1)$.  The inverse Compton
interaction of CMB photons with hot electrons in a galaxy cluster
modifies the CMB spectrum. For small optical depth this is well
described by
\begin{equation}
\delta I_{\rm SZ}(\nu) = i_0\,g(x)\,y,\;\mbox{with}\; y = 
\frac{\sigma_{\rm T}}{m_{\rm e}\,c^2}\,\int\!\! dl\, n_{\rm e}\,
kT_{\rm e},
\end{equation}
the Compton $y$-parameter.  $n_{\rm e}$, and $T_{\rm e}$ are the
electron density and temperature, respectively, $l$ is the coordinate
along the line-of-sight, and
\begin{equation}
\label{eq:g(x)}
g(x) = \frac{x^4\,e^x}{(e^x -1)^2} \left( x\,\frac{e^x +1}{e^x-1} -4
\right) \approx -2 \, x^2 \;\mbox{for}\; x \ll 1,
\end{equation}
describes the spectral distortions.\footnote{In the case of hot
clusters ($\sim 10$ keV) relativistic effects modify the shape of
$g(x)$. However, the corrections are negligible on the Rayleigh-Jeans
side of the CMB spectrum for $kT_{\rm e} \ll $ MeV \citep[e.g. see
Fig. 6 in][]{2000A&A...360..417E}, so that Eq. \ref{eq:g(x)} is an
accurate estimate in our frequency regime ($x\ll 1$). The relativistic
electron population in Coma is also not able to modify this behaviour
significantly
\citep{2000A&A...360..417E,2000ApJ...535L..71B,2002ApJ...575...12S}.}

The area integrated SZ distortion of the CMB by a galaxy cluster can
be described as a SZ-flux (even if this is negative in the
Rayleigh-Jeans part of the CMB):
\begin{equation}
F_{\rm SZ}(\nu) = \frac{2\,\sigma_{\rm T}\,i_0\,E_{\rm
th,e}}{3\,m_{\rm e}\,c^2\,D^2}\,g(x),
\end{equation}
where $E_{\rm th,e}$ is the total energy of all thermal cluster
electrons, and $D$ is the angular cluster distance.

In our examples, we assume the cluster to be isothermal, spherically
symmetric, and its electron density well described by a beta-model
$n_e(r) = n_{\rm e,0}\,[1+(r/r_{\rm c})^2]^{-3\,\beta/2}$ up to the
accretion shock wave located at radius $r=R$. For Coma, we adopt
$kT_{\rm e}= 8.2$ keV, $n_{\rm e,0} = 3\cdot 10^{-3}\,\h^{1/2} \, {\rm
cm^{-3}}$, $r_{\rm c} = 0.4\,\h^{-1}$ Mpc $ \hat{=} 10'$, $\beta =
0.8$ \citep{1992A&A...259L..31B}, $R= 5 \,\h^{-1} \,{\rm Mpc}$
\citep{1998AA...332..395E}, and $D= 140 \,\h^{-1} \,{\rm Mpc}$ . This
leads to $E_{\rm th,e}^{\rm Coma} = 7.7\cdot 10^{63}\,\h^{-5/2} \,{\rm
erg}$, and a SZ-luminosity of
\begin{equation}
\label{eq:szComa}
F_{\rm SZ}^{\rm Coma}(\nu) = 3.0\,g(x)\,\h^{-\frac{1}{2}}  {\rm Jy}
\approx -1.9\cdot 10^{-3}  \,\nu_{\rm GHz}^2 \h^{-\frac{1}{2}}  {\rm Jy}, 
\end{equation}
where $\nu_{\rm GHz} = \nu/{\rm GHz}$. The same parameters give a
central $y$-parameter of $y^{\rm Coma}_{\rm model} = 0.85 \cdot
10^{-4}\,\h^{-1/2}$, which is in good agreement with the measured
value of $y^{\rm Coma}_{\rm obs} = (0.7...1.0)\cdot 10^{-4}$
\citep{1995ApJ...449L...5H,2001ApJ...555L..11M,2002ApJ...574L.119D}.

\section{Coma radio halo}

\begin{figure}[t]
\begin{center}
\psfig{figure=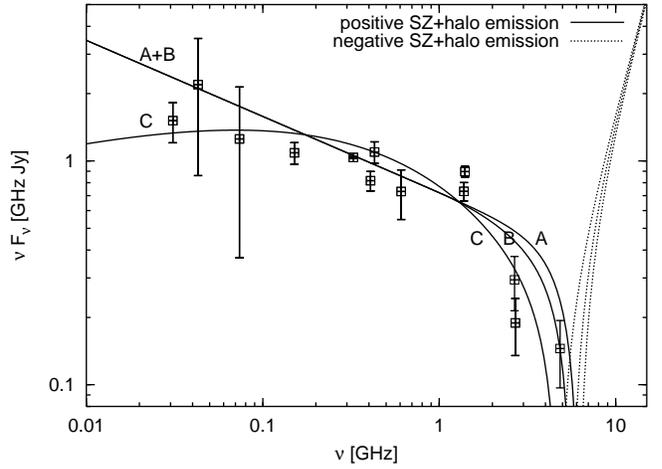,width=0.5 \textwidth,angle=0}
\end{center}
\vspace{-0.8cm}
\caption[]{\label{fig:spec2} Radio spectrum of the Coma cluster in
$\nu\,F(\nu)$ for better display (see Fig. \ref{fig:spec1}).  Solid
(dotted) curves represent the total cluster emission (CMB decrement)
of the three halo emission models (A,B,C) combined with the SZ
decrement.}
\end{figure}

Cluster radio halos have steep radio spectra ($F_{\rm halo} \propto
\nu^{-s}$) with spectral indexes $s >1$. In the case of Coma, there
seems to be a further steepening beyond 1.4 GHz. The current data as
compiled by \cite{astro-ph/0210147} is displayed in
Fig. \ref{fig:spec1} in comparison with three model spectra.\\ A) A
steep power law spectrum:
\begin{equation}
\label{eq:mod1}
F_{\rm halo,1}^{\rm Coma} (\nu) = 0.723\,\nu_{\rm GHz}^{-1.34} \,{\rm Jy}
\end{equation}
B) As A) but with a soft high frequency steepening:
\begin{equation}
\label{eq:mod2}
F_{\rm halo,2}^{\rm Coma} (\nu) = 0.723\,\nu_{\rm GHz}^{-1.34-0.1
\max(0,\ln \nu_{\rm GHz})}\,{\rm Jy}
\end{equation}
C) A flat power law spectrum with exponential cutoff:
\begin{equation}
\label{eq:mod3}
F_{\rm halo,3}^{\rm Coma} (\nu) = 3.42\, \nu_{\rm GHz}^{-0.8} \,
\exp(-\sqrt{\nu_{\rm GHz}})\,{\rm Jy}.
\end{equation}
The last spectrum is a fit to the data by \cite{astro-ph/0210147},
following a functional form proposed by \cite{1987A&A...182...21S} as
a model for a continously in-situ accelerated electron population with
exponential high energy cutoff. It should be noted that the functional
form of Eq. \ref{eq:mod3} was estimated with the monochromatic
approximation for the synchrotron emissivity of monoenergetic
electrons. With the correct calculations, therefore, the high
frequency cutoff would be much smoother.\footnote{The synchrotron
spectrum of a monoenergetic electron population in an isotropic
distribution of magnetic fields is displayed in
Fig. \ref{fig:spec1}. A synchrotron spectrum can not be more strongly
bent than this emission spectrum kernel.} In order to get a similar
sharp cutoff in the emission spectrum, a sharp cutoff in the electron
spectrum is required, as e.g. in the time dependent in-situ model
investigated by \cite{2001MNRAS.320..365B}.
Emission model C reproduces the observed data only if the SZ decrement
is neglected (see Fig. \ref{fig:spec2}). However, as explained above, this is not a
problem for the continuous in-situ model, since correctly estimated,
it produces a softer cutoff than given by model C.
In model C the SZ-effect would dominate above 4.7 GHz and lead to a
decrement at the observed frequency of 4.9 GHz in contrast to the
data. In model B (A) the change of sign of the total radio fluxes is
at 5.7 GHz (6.2 GHz). Model A predicts too much flux at high
frequencies, but model B seems to be consistent with the data. Thus,
only a weak spectral steepening of the radio halo emission above 1 GHz
is required by the data.

Such a weak steepening can also be explained in the secondary
electron model. There the radio halo spectrum steepening may occur due
to a steepening of the parent cosmic ray proton spectrum.

\section{Radial brightness profiles}

\begin{figure}[t]
\begin{center}
\psfig{figure=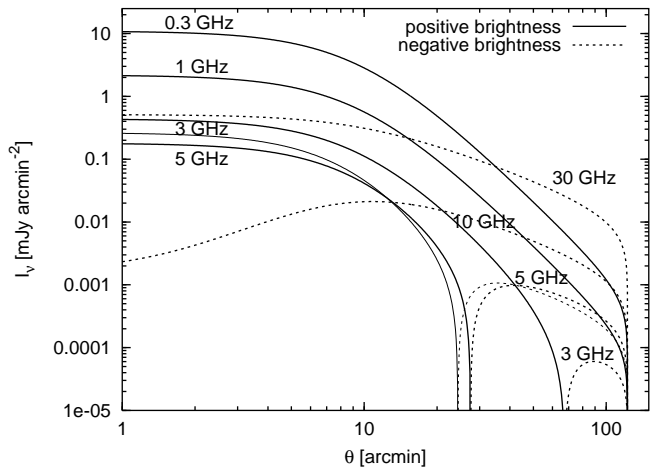,width=0.5 \textwidth,angle=0}
\end{center}
\vspace{-0.8cm}
\caption[]{\label{fig:prof} Radial brightness profile of our toy-model
of the Coma cluster at different frequencies as it results from
combining the radio halo emission and the SZ decrement profiles. The
total emissivity follows the curve of model B displayed in
Fig. \ref{fig:spec2}. At 0.3 GHz (30 GHz) the practically unmodified
radio halo (SZ decrement) profile is visible. The thin line
gives the 5 GHz model for $\alpha =1.1$, $\beta = 0.85$, illustrating the
strong dependence of the profiles on the parameters.}
\end{figure}

The various radio emission processes of galaxy clusters can partly be
separated by their morphological difference. The SZ-emission has a
spatially broad brightness distribution, since it is proportional (in
an isothermal model) to the line-of-sight projected electron
density. The radio emission seems to be as concentrated as the
Bremsstrahlung X-ray emission of the thermal cluster gas: $I_{\rm
halo} \propto I_{\rm X}^{\alpha}$, where $\alpha \approx 1$
\citep{2001A&A...369..441G}. Since the X-ray emissivity is
proportional to $n_{\rm e}^2$, we assume the radio halo emissivity to
exhibit the same scaling. Note, that secondary electron models would
usually predict a steeper scaling ($\alpha \approx 1.3$ by
\cite{2000A&A...362..151D}, but see \cite{2001ApJ...562..233M}), which
therefore seem to be disfavoured \citep{2001A&A...369..441G,
2002BrunettiTaiwan}.

In Fig. \ref{fig:prof} we used a toy model to illustrate the
consequences of the SZ decrement on the properties of the radio
brightness profiles as a function of frequency. For that purpose we
assume a uniform emission spectrum given by model B (Eq.
\ref{eq:mod2}) and the above mentioned radial scaling.\footnote{The
resulting spatially constant spectral index at low ($<$ GHz)
frequencies is not exactly what is observed since a radial spectral
steepening between 0.3 and 1.4 GHz is found by
\citep{1993ApJ...406..399G}.}. At low ($\lesssim$ GHz) frequencies the
radio halo emission dominates. With increasing frequency a negative
SZ-bowl surrounding the residual of the central visible radio halo
emission appears and grows (dotted line in Fig. \ref{fig:prof}). At
10 GHz the SZ decrement dominates at every angular position, but its
central part still reveals the presence of the radio halo emission by
exhibiting a dip in the SZ decrement. At 30 GHz the radio halo
emission is insignificant, so that this frequency can safely be used
for SZ-measurements.

Even at frequencies, at which the total radio emission is zero or
negative ($\gtrsim$ 5 GHz), a central positive emission region can
exist if the frequency is not too high ($<$ 10 GHz). Thus, a positive
flux may be attributed to such a radio map, if only the region with
positive flux is included into the map analysis. Especially, since the
negative bowl around the central region is relatively shallow (it
dominates only the total flux due to its larger area) there is always
the danger that it is regarded as noise, or as a remnant of imperfect
map reconstruction in the case of an interferometric measurement
without {\it zero-spacing flux}. In the case that the radio halo
luminosity of a cluster is assumed to be equal to the flux within the
positive region, the real halo luminosity would be underestimated
(e.g. by $34\%$ at 5 GHz in our toy-model).

In the frequency range of several GHz, there exist a relatively sharp
outer edge of the positive brightness region, which (in our toy-model)
is located at $\theta \approx 70'$ for 3 GHz, and at $\theta \approx
30'$ for 5 GHz, whereas the synchrotron emission extends up to $\theta
=120'$. The exact location of the edge of the positive emission region
depends sensitively on several of the model parameters ($r_{\rm c}$,
$\beta$, $\alpha$, see Fig. \ref{fig:prof}) and on here not modelled
details of the radio halo emission spectrum as a function of cluster
radius. However, a similar frequency trend of the edge of positive
emission can be seen by visual inspection of the Coma radio maps at
1.4 GHz \citep{1997A&A...321...55D} which exhibits positive emission
up to $\theta \gtrsim 45'$, and the maps at 2.7 and 4.9 GHz
\citep{astro-ph/0210147} which seem to be positive only up to $\theta
\approx 20'$ and $\theta \approx 10'$, respectively. Although there is
likely a contribution to this shrinking due to spectral steepening at
larger radii, our estimate and the actual SZ measurements argue that a
significant fraction of the emission area shrinking is due to the
unavoidable SZ-effect. The exact amount is difficult to estimate due
to our still limited knowledge of the exact radio halo emission
profile. However, the prediction of the SZ-decrement profile is quiet
robust for the central cluster regions due to the observationally well
constrained $y$-parameter with an error of only $\sim
10\%$. Therefore, it is possible to state that for the measurements of
\cite{astro-ph/0210147} at 2.7 GHz and 4.9 GHz the central
SZ-decrement is of the order of 30\% and 50\% of the 1-$\sigma$ noise
level (per beam as shown in their Fig. 2 and 3), respectively.  Since
the SZ decrement strength decreases radially only slowly, it should
affect the outer regions of the rapidly decreasing radio
halo\footnote{E.g. at the location of the 3-$\sigma$ brightness
contour, located at $10'$ ($7'$) in the 2.7 GHz (4.9 GHz) map of
\cite{astro-ph/0210147}, with a brightness of $0.038$ ($0.067$) ${\rm
mJy/arcmin^2}$, the negative SZ contribution is $-0.0026$ ($-0.011$)
${\rm mJy/arcmin^2}$.}. Thus it should be statistically detectable
there due to the sufficient large number of resolution elements
covering the cluster outskirts. A careful analysis of high
signal-to-noise high frequency maps of radio halos are expected to
reveal an extended decrement at outer cluster regions.

\section{Conclusion}

Measurements of cluster radio halos at several GHz are contaminated by
the thermal SZ effect, or vice versa, at $\sim 10$ GHz, measurements
of the SZ effect have to take the possible presence of cluster radio
halos into account. The morphology of the radio emission of a galaxy
cluster at such frequencies, where the total radio halo emission and
the total SZ decrement roughly compensate each other, is that of a
central emission region surrounded by a shallow negative bowl. This
complex morphology can lead to ambiguities in observational
determinations of radio halo fluxes, since the result depends on the
adopted procedure (flux from the full cluster area, or only
flux from regions of positive brightness). We therefore stress the
need to communicate up to which radius a cluster radio halo flux was
integrated. It would be best, if such measurements would be provided
for several radii. Radio halo emission measurements should carefully
be corrected for the unavoidable SZ decrement \citep[see
also][]{2000ApJ...544..686L}.

Using the Coma cluster as an example we demonstrated that the existing
radio maps at 2.7 and 4.9 GHz of the Coma radio halo should be
affected by the SZ decrement. This is very likely part of the reason
(in combination with radial spectral steepening of radio halo
emissivity and a decreasing S/N ratio) why the apparent radio halo
size decreases with increasing frequency. The observed sharp spectral
steepening of the total radio flux of Coma seems to be partly -- but
not completely -- due to the SZ effect. This implies that only a
weakly bended emission spectrum is required to explain the data, and
consequently the constrain arising from the apparent strong spectral
steepening on some proposed radio halo formation scenarios as
continuous in-situ acceleration and secondary electron injection is
relaxed.

\begin{acknowledgements}
I acknowledge several useful comments by E. Churazov, K. Jedamzik,
F. Miniati, and C. Vogt. This work
was done in the framework of the EC Research and Training
Network {\it The Cosmic Microwave Background}.
\end{acknowledgements}

%\appendix

%\bibliography{aamnem99,tae}

\begin{thebibliography}{}

\bibitem[\protect\astroncite{{B{\" o}hm} et~al.}{2002}]{astro-ph/0208458}
{B{\" o}hm}, C., {En{\ss}lin}, T.~A., {Silk}, J., 2002,
\newblock astro-ph/0208458

\bibitem[\protect\astroncite{{Blasi} et~al.}{2000}]{2000ApJ...535L..71B}
{Blasi}, P., {Olinto}, A.~V., {Stebbins}, A., 2000,
\newblock {\apjl} {535}, L71

\bibitem[\protect\astroncite{{Briel} et~al.}{1992}]{1992A&A...259L..31B}
{Briel}, U.~G., {Henry}, J.~P., {B{\"o}hringer}, H., 1992,
\newblock {\aap} {259}, L31

\bibitem[\protect\astroncite{{Brunetti}}{2002}]{2002BrunettiTaiwan}
{Brunetti}, G., 2002,
\newblock in S. {Bowyer} \& C.-Y. {Hwang} (eds.), {Matter and Energy in
  Clusters of Galaxies}, ASP Conf. Series,
\newblock in preparation, astro-ph/0208074

\bibitem[\protect\astroncite{{Brunetti} et~al.}{2001}]{2001MNRAS.320..365B}
{Brunetti}, G., {Setti}, G., {Feretti}, L., {Giovannini}, G., 2001,
\newblock {\mnras} {320}, 365

\bibitem[\protect\astroncite{{Buote}}{2001}]{2001ApJ...553L..15B}
{Buote}, D.~A., 2001,
\newblock {\apjl} {553}, L15

\bibitem[\protect\astroncite{{Colafrancesco} \&
  {Mele}}{2001}]{2001ApJ...562...24C}
{Colafrancesco}, S., {Mele}, B., 2001,
\newblock {\apj} {562}, 24

\bibitem[\protect\astroncite{{De Petris} et~al.}{2002}]{2002ApJ...574L.119D}
{De Petris}, M., {D'Alba}, L., {Lamagna}, L., et al., 2002,
\newblock {\apjl} {574}, L119

\bibitem[\protect\astroncite{{Deiss} et~al.}{1997}]{1997A&A...321...55D}
{Deiss}, B.~M., {Reich}, W., {Lesch}, H., {Wielebinski}, R., 1997,
\newblock {\aap} {321}, 55

\bibitem[\protect\astroncite{{Dennison}}{1980}]{1980ApJ...239L..93D}
{Dennison}, B., 1980,
\newblock {\apjl} {239}, L93

\bibitem[\protect\astroncite{{Dolag} \&
  {En{\ss}lin}}{2000}]{2000A&A...362..151D}
{Dolag}, K., {En{\ss}lin}, T.~A., 2000,
\newblock {\aap} {362}, 151

\bibitem[\protect\astroncite{{En{\ss}lin} et~al.}{1998}]{1998AA...332..395E}
{En{\ss}lin}, T.~A., {Biermann}, P.~L., {Klein}, U., {Kohle}, S., 1998,
\newblock {\aap} {332}, 395

\bibitem[\protect\astroncite{{En{\ss}lin} \&
  {Kaiser}}{2000}]{2000A&A...360..417E}
{En{\ss}lin}, T.~A., {Kaiser}, C.~R., 2000,
\newblock {\aap} {360}, 417

\bibitem[\protect\astroncite{{En{\ss}lin} \&
  {R{\"o}ttgering}}{2002}]{astro-ph/0209218}
{En{\ss}lin}, T.~A., {R{\"o}ttgering}, H., 2002,
\newblock {\aap} in press,
\newblock astro-ph/0209218

\bibitem[\protect\astroncite{{Giovannini} \&
  {Feretti}}{2000}]{2000NewA....5..335G}
{Giovannini}, G., {Feretti}, L., 2000,
\newblock {New Astronomy} {5}, 335

\bibitem[\protect\astroncite{{Giovannini} et~al.}{1993}]{1993ApJ...406..399G}
{Giovannini}, G., {Feretti}, L., {Venturi}, T., {Kim}, K.~T., {Kronberg},
  P.~P., 1993,
\newblock {\apj} {406}, 399

\bibitem[\protect\astroncite{{Giovannini} et~al.}{1999}]{1999NewA....4..141G}
{Giovannini}, G., {Tordi}, M., {Feretti}, L., 1999,
\newblock {New Astronomy} {4}, 141

\bibitem[\protect\astroncite{{Govoni} et~al.}{2001}]{2001A&A...369..441G}
{Govoni}, F., {En{\ss}lin}, T.~A., {Feretti}, L., {Giovannini}, G., 2001,
\newblock {\aap} {369}, 441

\bibitem[\protect\astroncite{{Herbig} et~al.}{1995}]{1995ApJ...449L...5H}
{Herbig}, T., {Lawrence}, C.~R., {Readhead}, A.~C.~S., {Gulkis}, S., 1995,
\newblock {\apjl} {449}, L5

\bibitem[\protect\astroncite{{Jaffe}}{1977}]{1977ApJ...212....1J}
{Jaffe}, W.~J., 1977,
\newblock {\apj} {212}, 1

\bibitem[\protect\astroncite{{Kempner} \&
  {Sarazin}}{2001}]{2001ApJ...548..639K}
{Kempner}, J.~C., {Sarazin}, C.~L., 2001,
\newblock {\apj} {548}, 639

\bibitem[\protect\astroncite{{Liang}
et~al.}{2000}]{2000ApJ...544..686L} Liang, H., Hunstead, R.~W.,
Birkinshaw, M., Andreani, P.,2000, \newblock {\apj} {544}, 686

\bibitem[\protect\astroncite{{Mason} et~al.}{2001}]{2001ApJ...555L..11M}
{Mason}, B.~S., {Myers}, S.~T., {Readhead}, A.~C.~S., 2001,
\newblock {\apjl} {555}, L11


\bibitem[\protect\astroncite{{Miniati} et~al.}{2001}]{2001ApJ...562..233M} 
Miniati, F., Jones, T.~W., Kang,
H., Ryu, D., 2001, \newblock {\apj} {562}, 233


\bibitem[\protect\astroncite{{Schlickeiser} et~al.}{1987}]{1987A&A...182...21S}
{Schlickeiser}, R., {Sievers}, A., {Thiemann}, H., 1987,
\newblock {\aap} {182}, 21

\bibitem[\protect\astroncite{{Shimon} \&
  {Rephaeli}}{2002}]{2002ApJ...575...12S}
{Shimon}, M., {Rephaeli}, Y., 2002,
\newblock {\apj} {575}, 12

\bibitem[\protect\astroncite{{Sunyaev} \&
  {Zeldovich}}{1980}]{1980ARA&A..18..537S}
{Sunyaev}, R.~A., {Zeldovich}, I.~B., 1980,
\newblock {\araa} {18}, 537

\bibitem[\protect\astroncite{{Thierbach} et~al.}{2002}]{astro-ph/0210147}
{Thierbach}, M., {Klein}, U., {Wielebinski}, R., 2002,
\newblock {\aap} in press,
\newblock astro-ph/0210147

\bibitem[\protect\astroncite{{Zeldovich} \&
  {Sunyaev}}{1969}]{1969Ap&SS...4..301Z}
{Zeldovich}, Y.~B., {Sunyaev}, R.~A., 1969,
\newblock {\apss} {4}, 301

\end{thebibliography}
\bibliographystyle{aabib99}

\end{document}